\def\papertitle{Transformer-Based Plate Parameter Estimation with Differentiable and Particle-Swarm Refinement}
\def\paperauthorA{David Marttila}
\def\paperauthorB{Rodrigo Diaz}
\def\paperauthorC{Pablo Tablas De Paula}
\def\paperauthorD{Ilias Ibnyahya}
\def\paperauthorE{Chin-Yun Yu}
\documentclass[twoside,a4paper]{article}
\usepackage{etoolbox}
\usepackage[taskA]{dafx26challenge}
\usepackage{amsmath,amssymb,amsfonts,amsthm}
\usepackage{siunitx}
\usepackage{euscript}
\usepackage[T1]{fontenc}
\usepackage[utf8]{inputenc}
\usepackage{iftex}
\usepackage[english]{babel}
\usepackage{caption}
\usepackage{subfig}
\usepackage{booktabs}
\usepackage{color}
\usepackage{graphicx}
\usepackage{url}
\ifPDFTeX
  \input glyphtounicode
\fi
\ninept

\newcounter{numauth}
\newcounter{listcnt}
\newcommand\authcnt[1]{\ifdefined#1 \stepcounter{numauth} \fi}
\newcommand\addauth[1]{
  \ifdefined#1
    \stepcounter{listcnt}
    \ifnum \value{listcnt}<\value{numauth}
      \appto\authorslist{, #1}
    \else
      \appto\authorslist{~and~#1}
    \fi
  \fi
}
\authcnt{\paperauthorB}
\authcnt{\paperauthorC}
\authcnt{\paperauthorD}
\authcnt{\paperauthorE}
\authcnt{\paperauthorF}
\authcnt{\paperauthorG}
\authcnt{\paperauthorH}
\authcnt{\paperauthorI}
\authcnt{\paperauthorJ}
\def\authorslist{\paperauthorA}
\addauth{\paperauthorB}
\addauth{\paperauthorC}
\addauth{\paperauthorD}
\addauth{\paperauthorE}
\addauth{\paperauthorF}
\addauth{\paperauthorG}
\addauth{\paperauthorH}
\addauth{\paperauthorI}
\addauth{\paperauthorJ}

\usepackage{times}
\usepackage[
  pdftitle={\papertitle},
  pdfauthor={\authorslist},
  pdfsubject={Proceedings of the 29th International Conference on Digital Audio Effects (DAFx26)},
  colorlinks=false,
  bookmarksnumbered,
  pdfstartview=XYZ]{hyperref}
\ifPDFTeX
\fi

\newcommand{\code}[1]{\begingroup\urlstyle{tt}\nolinkurl{#1}\endgroup}
\title{\papertitle}
\affiliation{\paperauthorA\textsuperscript{1},
\paperauthorB\textsuperscript{1,2}, \paperauthorC\textsuperscript{1},\\
\paperauthorD\textsuperscript{1}, and \paperauthorE\textsuperscript{1}}
{\textsuperscript{1}Queen Mary University of London, London, United Kingdom\\
\textsuperscript{2}Fraunhofer HHI, Berlin, Germany}
\begin{document}
\maketitle
\begin{abstract}
We present two Transformer-based methods for Task A of the 1st DAFx Parameter
Estimation Challenge, which requires estimating six effective physical
parameters of a synthetic plate-reverb model from its impulse response (IR).
Method A1 combines an Audio Spectrogram Transformer encoder and Transformer
regressor with differentiable IR refinement. Method A2 uses the same encoder to
condition a continuous normalizing flow and refines sampled candidates using
particle swarm optimisation (PSO) and gradient polishing. Both methods preserve
the absolute IR scale to recover surface density. On a synthetic holdout set of
100 IRs, both refinement procedures reduce waveform and parameter errors by
more than three orders of magnitude relative to the unrefined neural outputs.
The PSO-based pipeline achieves the lowest errors, indicating near-perfect
recovery in this matched synthetic setting.
\end{abstract}

\section{Introduction}
\label{sec:introduction}
Task A of the challenge asks participants to identify parameters of a damped
rectangular plate from measured impulse responses (IRs) generated by a
reference simulator~\cite{Ducceschi:2026:Challenge}. The effective physical
parameter vector is
\begin{equation}
  \theta = \{\mu, D/\mu, T_0/\mu, L_y, x_o, y_o\},
  \label{eq:task-a-vector}
\end{equation}
where $\mu=\rho h$ is surface density. Both methods use the unnormalised
displacement IR from the provided \code{.npz} files when available.
Peak-normalised \code{.wav} renderings are treated as listening checks only,
because they discard the absolute amplitude information needed for $\mu$.
The inverse methods do not use the closed-form plate modal distribution or
frequency-dependent damping law as direct side information, following the
challenge restriction.

Our implementation, including data generation, model training, and parameter
refinement, is available as open-source code.\footnote{\href{https://github.com/davidmarttila/dafx_challenge}{\nolinkurl{github.com/davidmarttila/dafx_challenge}}}

\section{Training Data}
\label{sec:training-data}
Training data was generated with the reference plate model and the challenge
parametrisation for the six identifiable physical parameters. Using a
GPU-accelerated implementation, we generated 20 HDF5 shards. With the
generator defaults, each shard contains $512 \cdot 32 = 16{,}384$ five-second IRs at
\SI{44.1}{\kilo\hertz}, giving $327{,}680$ IRs in total.

Each example contains a peak-normalised IR and its pre-normalisation peak
$g_x=\max_n |x[n]|$. It also contains normalised values for the six physical
parameters $\mu$, $D/\mu$, $T_0/\mu$, $L_y$, $x_o$, and $y_o$. Retaining $g_x$
preserves the absolute displacement scale used to recover $\mu$. Before
feature extraction, we downsample each IR to \SI{22.05}{\kilo\hertz}, since
the synthetic IRs contain no modes above \SI{10}{\kilo\hertz}.

\section{Task A: Physical Parameters}
\label{sec:task-a}
Task A submissions produce one CSV estimate per target IR with the six columns
\code{mu}, \code{D_mu}, \code{T0_mu}, \code{Ly}, \code{op_x}, and
\code{op_y}. They are trained in a supervised manner to take an IR as input and predict the associated control parameters, normalised to the $[-1, 1]$ range.

\subsection{Input Features}

After downsampling, the peak-normalised IR is passed through an encoder inspired
by the Audio Spectrogram Transformer (AST)~\cite{Gong:2021:AST}. Rather than
using a single STFT, we extract three feature representations from the IR, each
of which is fed to its own convolutional neural network (CNN) stem:

\begin{enumerate}
    \item The log-magnitude spectrum of the entire (decimated) IR
    \item The time-domain waveform, cropped to a length of 1 second, to preserve phase and transient behaviour
    \item An STFT of size 4096 with hop length 1024, whose frequency bins are summarised into 64 bands to show the overall decay behaviour
\end{enumerate}

Each of these stems produces feature vectors, to which positional embeddings are added. We linearly embed the IR's pre-normalisation global gain as an additional feature. In total, we obtain 148 feature vectors of size 256.

\begin{figure*}[t]
  \centering
  \includegraphics[width=\textwidth]{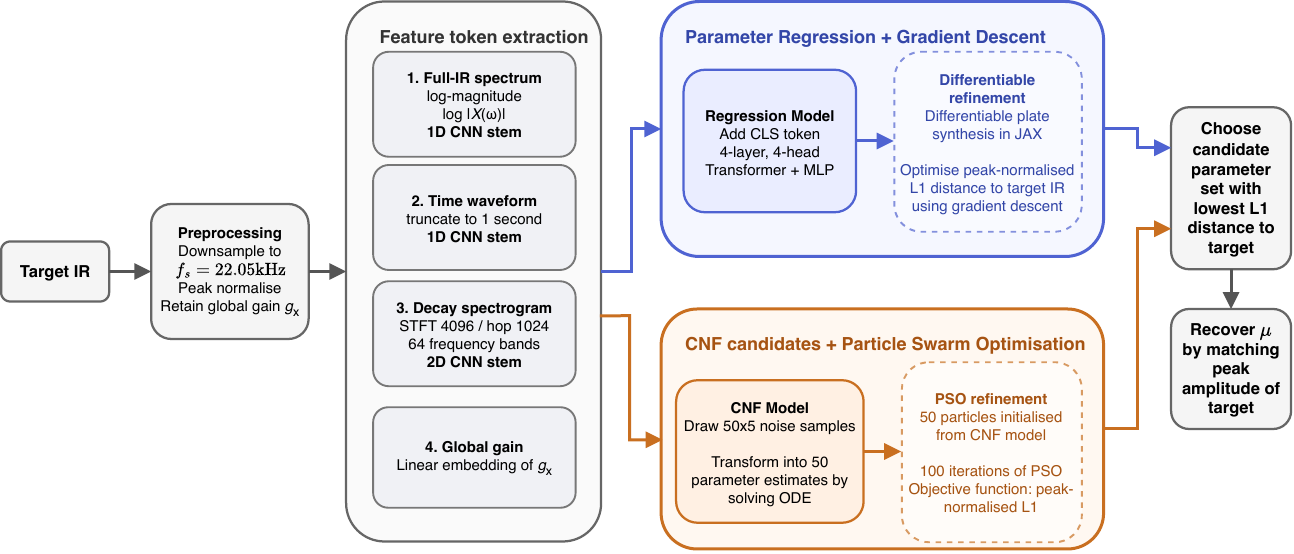}
  \caption{Task A estimation pipelines. Blue denotes Method A1, and orange
  denotes Method A2. A shared encoder forms feature tokens from the full-IR log
  spectrum, one-second waveform, decay spectrogram, and
  retained global gain. The regression model refines its point estimate by
  differentiating through the plate synthesiser, whereas the conditional flow
  model supplies candidates for particle swarm optimisation (PSO). The
  lowest-error candidate is selected, and the surface density $\mu$ is recovered
  from the absolute peak amplitude.}
  \label{fig:task-a-overview}
\end{figure*}

Figure~\ref{fig:task-a-overview} summarises the shared feature-token extraction
and the two inference pipelines, including their respective refinements and
common recovery of the surface density.

\subsection{Method A1: Regression + Differentiable Refinement}
\label{ssec:task-a-transformer}
In the first method, we train a transformer to predict the ground truth parameters directly from the input features. As in the standard AST architecture, we add an additional CLS token to the input features. The resulting 149 vectors are processed by a transformer with 4 layers, each containing 4 attention heads. After the final layer, a feed-forward multi-layer perceptron (MLP) predicts the parameters from the CLS token. The full network contains 2.9M trainable parameters.

We train the network to predict the five parameters $D/\mu$, $T_0/\mu$, $L_y$, $x_o$ and $y_o$ in a supervised manner, using the normalised squared error as the training objective. We train the network for 200 epochs, taking about 20 hours on a single A100 GPU, and select the checkpoint with the lowest NMSE on a held-out validation partition of the synthetic data.

During inference, to recover the control parameters for the given test IRs, we further refine the network's predictions using gradient descent. We implement the modal IR synthesis in a differentiable manner using JAX, and use the Adam optimiser with a learning rate of $10^{-4}$ to minimise the L1 waveform loss between the target IR and the IR associated with the current predicted parameters. Both IRs are peak-normalised for loss calculation.

For a large number of modes, backpropagating through the synthesis of a full five-second IR can require too much memory for a single GPU. To combat this, we limit the synthesis and distance calculations to only the first \SI{500}{\milli\second}. We run the optimisation for 100 steps, taking about 5 minutes per IR on a single A100 GPU.

Since $\mu$ directly relates to the overall amplitude of the IR \cite{Ducceschi:2026:Challenge}, we do not train the network to predict it. Instead, we first predict and optimise the other five parameters using the process above. Because the refinement minimises the peak-normalised error, $\mu$ is then recovered analytically by matching the peak gain of the synthesised IR to that of the target.

\begin{table*}[t]
  \centering
  \begin{minipage}[t]{0.48\textwidth}
    \centering
    \captionof{table}{Regression-network performance on the 100-IR synthetic
    holdout set, before and after differentiable refinement.}
    \label{tab:regression-refinement}
    \footnotesize
    \setlength{\tabcolsep}{4pt}
    \begin{tabular}{@{}lcc@{}}
      \toprule
      Method &
      \shortstack{Peak-normalised\\waveform L1} &
      \shortstack{Parameter\\NMSE} \\
      \midrule
      Regression model
        & $1.13 \cdot 10^{-2}$
        & $1.47 \cdot 10^{-4}$ \\
      Regression + refinement
        & $9.36 \cdot 10^{-6}$
        & $8.72 \cdot 10^{-8}$ \\
      \bottomrule
    \end{tabular}
  \end{minipage}
  \hfill
  \begin{minipage}[t]{0.48\textwidth}
    \centering
    \captionof{table}{Flow-network performance on the 100-IR synthetic holdout
    set, before and after PSO and gradient polishing.}
    \label{tab:flow-sampling-refinement}
    \footnotesize
    \setlength{\tabcolsep}{4pt}
    \begin{tabular}{@{}lcc@{}}
      \toprule
      Method &
      \shortstack{Peak-normalised\\waveform L1} &
      \shortstack{Parameter\\NMSE} \\
      \midrule
      Best of 50 flow samples
        & $2.13 \cdot 10^{-2}$
        & $4.81 \cdot 10^{-5}$ \\
      PSO + gradient polishing
        & $6.84 \cdot 10^{-6}$
        & $4.73 \cdot 10^{-10}$ \\
      \bottomrule
    \end{tabular}
  \end{minipage}
\end{table*}

\subsection{Method A2: Continuous Normalizing Flow + PSO}
\label{ssec:task-a-flow}
The second Task A candidate uses the same AST encoder as Method~A1.
Instead of estimating the parameters directly through regression, however, we
use a continuous normalizing flow (CNF) to map samples from a base distribution
to conditional parameter samples.

We train a Diffusion Transformer (DiT) using rectified flow matching. Given a timestep $\tau\in[0, 1]$, the AST conditioning output, and sampled noise, we linearly interpolate between the noise and the ground-truth parameters and train the DiT to predict the corresponding vector field at $\tau$. The DiT has 4 layers with 4 heads each. Its design is based on the audio-conditioned parameter-estimator architecture of Hayes et al.~\cite{Hayes:2025:SynthInversion}. We do not use their
permutation-invariant or equivariant components because the Task A
parameterisation has no relevant symmetries.

The AST encoder and DiT together contain 6.5M trainable parameters. We train the networks jointly for 100 epochs, which takes about 45 hours on a single A100 GPU.

Like the regression architecture, the flow matching approach predicts five parameters excluding $\mu$. During inference, we sample 50 five-dimensional vectors of uniform noise and solve the ODE given by the DiT using a Tsit5 solver with 30 timesteps. This yields 50 samples from the predicted parameter distribution. We select the checkpoint with the lowest average NMSE across the 50 samples, calculated on a held-out validation partition of the synthetic data.

To match the target IRs, we use the 50 samples to initialise a particle swarm optimisation (PSO) refinement. As in the differentiable approach, this gradient-free refinement minimises the peak-normalised L1 waveform error. After optimisation, we recover $\mu$ by matching the peak gain of the best candidate's synthesised IR to that of the target IR. By synthesising the IRs using the same GPU-accelerated implementation that was used to generate the synthetic dataset, we can run PSO for 100 iterations in about three minutes.

\subsection{Evaluation on a Synthetic Holdout Set}
\label{ssec:task-a-evaluation}
Following the challenge notation, Task A uses NMSE over the six physical
parameters ($N_S=6$):
\begin{equation}
  \mathrm{NMSE}
  = \frac{1}{N_S}\sum_{i=1}^{N_S}
  \left(
    \frac{\alpha_i^{(\mathrm{est})}-\alpha_i^{(\mathrm{ref})}}
         {\alpha_i^{(\max)}-\alpha_i^{(\min)}}
  \right)^2,
  \label{eq:task-a-nmse}
\end{equation}
Tables~\ref{tab:regression-refinement} and
\ref{tab:flow-sampling-refinement} compare each neural estimator alone with its
respective refinement on a synthetic holdout set of 100 IRs.

For Method A1, the unrefined result is the Transformer's direct point estimate;
the second row applies differentiable gradient-based refinement. For Method A2,
the unrefined result is the best of 50 samples drawn from the conditional flow.
Because the lowest-error sample is selected independently for each metric, this
row is an oracle diagnostic of the unrefined sample set rather than a deployable
selection rule. The second flow row uses these samples to initialise PSO and
then applies a final gradient-polishing step. Differentiable refinement reduces
both errors by more than three orders of magnitude, while the PSO pipeline
reduces waveform error by more than three orders and parameter NMSE by
approximately five orders of magnitude.

\section{Conclusion}
\label{sec:conclusion}
We presented two methods for physical-parameter estimation in Task A. Method A1
produces a point estimate and refines five parameters by differentiating through
the plate synthesiser. Method A2 learns a conditional parameter distribution
and refines sampled candidates using particle swarm optimisation. Both methods
share an amplitude-aware encoder and recover surface density analytically from
the absolute peak gain after refinement. On the 100-IR synthetic holdout set,
both refinement procedures yielded large improvements over their respective
neural estimators alone. The PSO-based pipeline was substantially stronger
overall: after gradient polishing, it achieved the lowest waveform error and a
parameter NMSE more than two orders of magnitude below that of the
differentiably refined regression model. This amounts to near-perfect parameter
recovery in the matched synthetic scenario, although performance on measured or
otherwise mismatched IRs remains to be established.

\bibliographystyle{IEEEtranDAFx}
\bibliography{references}
\end{document}